\documentclass[%
reprint,
amsmath,amssymb,
aps, superscriptaddress
]{revtex4-2}
\usepackage{natbib}
\usepackage{graphicx}
\usepackage{dcolumn}
\usepackage{bm}
\usepackage{color}
\begin{document}

\title{Beyond the Cavity: \\ Molecular Strong Coupling using an Open Fabry-Perot Cavity}

\author{K. S. Menghrajani}
\email{kishan.menghrajani@monash.edu}
\affiliation{Department of Physics and Astronomy, University of Exeter, Exeter, EX4 4QL, United Kingdom
}
\affiliation{School of Physics and Astronomy, Faculty of Science, Monash University, Melbourne, Victoria 3800, Australia}

\author{B. J. Bower}
\affiliation{Department of Chemistry, University of Sheffield, Brook Hill, Sheffield, S3 7HF, United Kingdom}

\author{G. J. Leggett}
\affiliation{Department of Chemistry, University of Sheffield, Brook Hill, Sheffield, S3 7HF, United Kingdom}

\author{W. L. Barnes}
\email{w.l.barnes@exeter.ac.uk}
\affiliation{Department of Physics and Astronomy, University of Exeter, Exeter, EX4 4QL, United Kingdom
}%

\begin{abstract}
The coherent strong coupling of molecules with confined light fields to create polaritons -- part matter, part light -- is opening exciting opportunities ranging from extended exciton transport and inter-molecular energy transfer to modified chemistry and material properties. In many of the envisaged applications open access to the molecules involved is vital, as is independent control over polariton dispersion, and spatial uniformity. Existing cavity designs are not able to offer all of these advantages simultaneously. Here we demonstrate an alternative yet simple cavity design that exhibits all of the the desired features. We hope the approach we offer here will provide a new technology platform to both study and exploit molecular strong coupling. Although our experimental demonstration is based on excitonic strong coupling, we also indicate how the approach might also be achieved for vibrational strong coupling. 
\end{abstract}


\maketitle
\newpage


\section{\label{sec:intor}Introduction}

Molecular strong coupling has emerged as a new paradigm in nanophotonics that bridges physics, chemistry and materials science. When a large number of molecular vibronic or excitonic resonators are located within the confined light field associated with a cavity mode the molecules and cavity mode may interact. For low values of this interaction (low coupling strength) the emission and absorption of light may be modified, but the underlying molecular resonances remain unchanged; this is the weak coupling regime, important in areas such as single photon generation. For higher values of the coupling strength the interaction leads to the formation of hybridised (polariton) states, states that are part light and part matter. Strong coupling occurs when the coupling rate between the molecular resonances and the cavity mode exceeds the dissipation rates. There is intense activity in the field of strong coupling, particularly owing to the prospect of using vibrational strong coupling to modify a range chemical reactions~\cite{Ebbesen_ACS_Accounts_2016_49_2403,Herrera_JCP_2020_152_100902}. Whilst most research is currently focused, quite rightly, on trying to understand the complex coherent molecular process involved, the role played by the cavity mode is vital but much less explored. In this report we introduce a powerful new platform with which to pursue molecular strong coupling that we hope will find  applicability in many areas.

For polariton chemistry investigations, and for associated applications, we can identify a number of key attributes that any experimental cavity design should embody, they are: {First, open access:} we would like the strongly coupled molecules to be `open access', i.e. we want to be able to easily introduce and carry away reactants and products. {Second, dispersion:} we need appropriate control over the dispersion of the polaritons. It has become clear that not all `cavity' modes offer dispersion properties suitable for polariton chemistry. In particular it seems to be increasingly well-established that a high density of polariton states around zero momentum (the $k=0$ condition) are needed~\cite{Simpkins_JPCC_2021_125_19081}. {Third, spatial homogeneity:} we would prefer to have a system where all molecules involved are coupled to the same extent with the cavity mode, i.e. we seek spatial uniformity. {Fourth, independent tuning:} we would like to be able -- for example -- to modify the concentration of the molecular material without at the same time changing the extent of any de-tuning, i.e. the mismatch between the molecular resonance energy and the cavity mode energy. As far as we are aware, no existing cavity design offers all of these attributes at the same time. 

Most studies of molecular strong coupling have focused on planar (Fabry-Perot) optical microcavities where molecules are positioned between two closely spaced metal or dielectric mirrors. However, these structures offer only limited access to the molecules involved, making their use in cavity-modified chemistry possible, but difficult~\cite{Ahn_Science_2023_380_6650}. To overcome this limitation, alternative `open' geometries have been investigated, including surface plasmon modes~\cite{Baieva_JCP_2013_138_044707,Torma_RepProgPhys_2015_78_013901}, dielectric microspheres~\cite{Vasista_NL_2020_20_1766}, and surface lattice resonances~\cite{Yadav_NL_2020_20_5043,Verdelli_JPCC_2022_126_7143}. These geometries are very interesting, but the extent of the light-matter coupling now varies spatially across the structure, again making their use in cavity-modified chemistry challenging. Recently, another class of `cavity-free' geometry has been explored. These structures show extensive mode splitting without employing metallic or dielectric multi-layer mirrors (DBR); instead these geometries rely on reflection from the interface between the molecular material and another dielectric, such as air, to generate optical modes~\cite{Georgiou_JPCL_2020_11_9893,Thomas_JPCL_2021_12_6914,Canales_JCP_2021_154_024701}. Although changes in molecular absorption have been observed in some of these structures, their effectiveness in controlling chemistry is not yet established. Even if this type of open cavity can be convincingly put into the strong coupling regime, there is one problem they do not seem likely to help us overcome, that of achieving appropriate modal dispersion. Similarly, surface plasmons on planar metal films~\cite{Torma_RepProgPhys_2015_78_013901}, a system that otherwise looks very attractive, does not afford the appropriate dispersion. None of the cavity designs discussed thus far provide all of the desirable features we seek, see table 1. 

\begin{figure}[h!]
    \centering
    \includegraphics[width=0.95\linewidth]{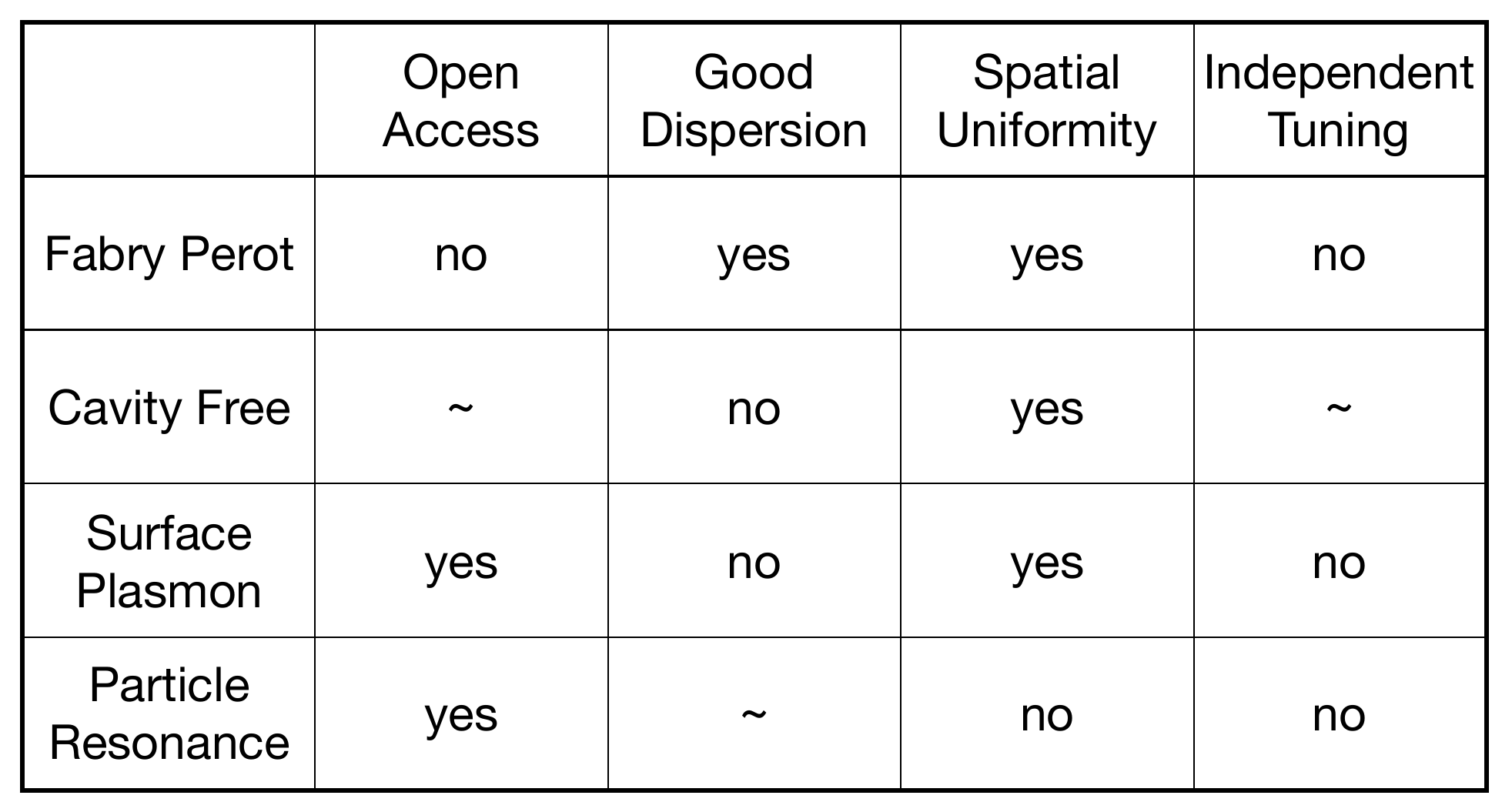}
\caption{\textbf{Table 1}  Summary of different cavity types commonly employed in strong coupling experiments, and attributes valuable for polaritonic chemistry.}
\label{fig:table 1}
\end{figure}

The strong coupling geometry we discuss here exploits the appropriate dispersion control \textbf{\textit{and}} the spatial uniformity offered by planar Fabry-Perot cavities, yet overcomes the lack of open access to the coupled molecules by placing the molecules outside rather than inside the optical cavity. In doing so we couple the molecules to the cavity mode via the evanescent (near) field of the cavity modes. Evanescent wave coupling is a well established approach in such areas as integrated optic couplers~\cite{Ahn_JLT_2010_28_3387}, lasers~\cite{Moon_APL_2004_84_4547} and amplifiers~\cite{Fang_MatToday_2007_10_28}.

Here we demonstrate this new approach using two different molecular systems. In the first we employ the well known J-aggregated dye TDBC~\cite{Balasubrahmaniyam_NatMat_2023_22_338}. This dye has been used for a large number of strong coupling experiments owing to the narrow bandwidth and high oscillator strength of the aggregates~\cite{Dintinger_PRB_2005_71_035424,Hobson_APL_2002_81_3519}. We used TDBC to compare the `standard' arrangement where the TDBC molecules are located inside the cavity (figure \ref{fig:schematics_and_PL}, left schematic) with our `new' arrangement where the TDBC molecules are outside the cavity, i.e. located in the evanescent tail of the cavity mode, (figure \ref{fig:schematics_and_PL}, middle schematic). We also employ a different dye, Rhodamine-B (RhB), for our outside the cavity arrangement, (figure \ref{fig:schematics_and_PL}, right schematic); RhB has a much broader emission spectrum than the narrow spectra offered by aggregated dyes such as TDBC. We make use of a combination of photoluminescence dispersion data, reflectance dispersion data and results from both numerical modelling and a simple coupled oscillator model. We especially focus on photoluminescence as this has proven to be a stronger measure of strong coupling compared to, for example, reflectance and  transmittance~\cite{Wersall_ACSPhot_2019_6_2570}. Our cavities were based on two gold (40 nm) mirrors separated by a PMMA spacer layer.

\section*{Results and Discussion}

\begin{figure*}[t!]
    \centering
    \includegraphics[width=1.00\linewidth]{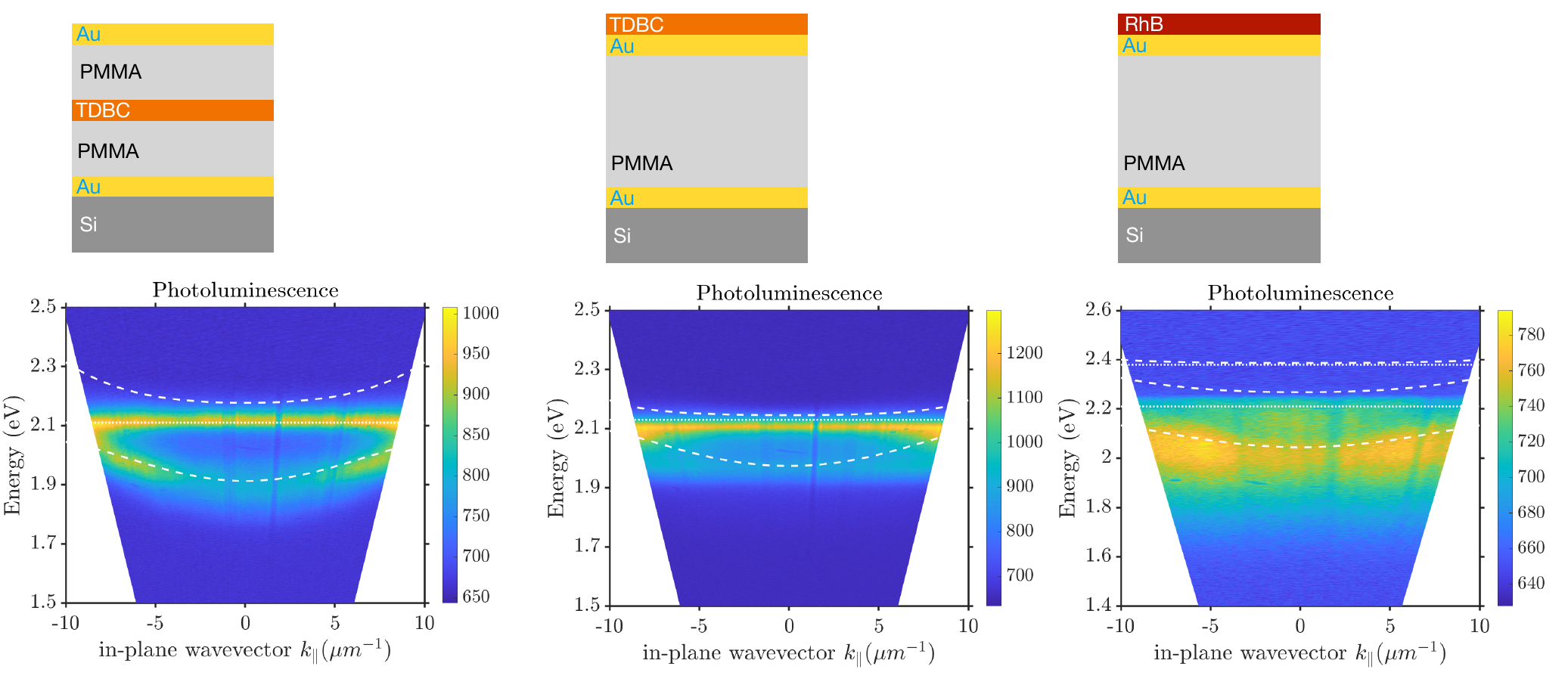}
\caption{\textbf{Sample Schematics and PL Dispersion.} \textbf{Upper row:} The three different cavities explored are shown as schematics. The left-hand two were used with TDBC films, the right-hand on with Rhodamine-B. \textbf{Lower row:} Photoluminescence data from the three samples shown as dispersion maps, i.e. the PL is shown on a colour scale as a function of frequency and in-plane wavevector, the plane corresponding to that of the cavities. Also shown as white dashed lines in each dispersion plot are the hybrid modes as determined by matching a coupled oscillator model to the PL and reflectance data.}
\label{fig:schematics_and_PL}
\end{figure*}

Schematics of the different structures we investigated are shown in the upper row of figure \ref{fig:schematics_and_PL}, in the lower row the corresponding dispersion diagrams based on measured photoluminescence data are shown. Reflectance data are shown in the SI.\\

\noindent\textbf{TDBC inside the cavity.}
In the left-hand column, upper panel, we show the familiar situation for strong coupling using a planar Fabry-Perot cavity, the dye-doped layer lies inside the cavity. The dye, here TDBC, was placed between two layers of PMMA using a layer-by-layer approach to deposit 4 monolayers (8 nm thick in total), see Vasista \textit{et al.} for details~\cite{Vasista_Nanoscale_2021_13_14497}. 
In the lower panel we show a dispersion plot based on the photoluminescence collected from the sample, see supplementary information. As expected, the PL shows emission mediated by the lower polariton, the PL peak tracking the dispersion of the lower polariton.

In addition we collected reflectance data, again see the supplementary information. We matched the results from a simple coupled oscillator model (see SI) to the reflectance and PL data, the positions for the polaritons obtained by simultaneously trying to match the model to the PL and reflectance data are indicated as dashed white lines on the PL dispersion. We can see that the PL nicely tracks the lower polariton, as expected, even though there are only 4 monolayers of TDBC within the cavity~\cite{Vasista_Nanoscale_2021_13_14497}.\\

At this stage it is worth looking at the numbers involved. A standard strong coupling criterion is~\cite{Rider_CP_2021_62_217,Heilmann_Nanophot_2020_9_267} that the Rabi splitting ($\Omega_R$) should exceed the average of the cavity (K) and molecular transition ($\Gamma$) linewidths, i.e. that,

\begin{align}
\Omega_R > (K+\Gamma)/2.
\label{eq:sc_condition}
\end{align}

\noindent For TDBC we find the molecular resonance linewidth is $\Gamma = 0.08\,\pm 0.01$ eV, whilst the bare cavity mode decay rate (see SI) is found to be $K = 0.11\,\pm 0.01$ eV.  From our match of a coupled oscillator model to our PL and reflectance data we found $\Omega_R=0.15\,\pm 0.03$ eV, so that in this case $\Omega_R/(K+\Gamma)/2$ = 1.5 $\pm$ 0.3.  Consequently, for this TDBC inside the cavity sample we meet the strong coupling condition.\\

\noindent\textbf{TDBC outside cavity.}
For the centre column the TDBC dye layer is no longer inside the cavity, rather it is now just outside the cavity, the TDBC molecules are thus openly accessible, see upper panel. In the lower panel we again show the PL dispersion data. Although the PL no longer tracks the lower polariton, there is clear evidence of significant coupling, the PL spectrum being very significantly modified when compared to PL from a thin TDBC film, see SI. This is a remarkable finding, there is no cavity between the dye and the detector, so the modified PL can not be the result of a simple cavity filtering effect. From our match of a coupled oscillator model to both our PL and reflectance data for this system we find that the Rabi splitting is $\Omega_R$=0.10 $\pm 0.02$ eV, and that in this case, the cavity linewidth is $K=0.18\,\pm 0.02$ eV. For our TDBC outside the cavity case we thus have $\Omega_R/(K+\Gamma)/2$ = 0.7 $\pm$ 0.3. Despite being -- at best -- just at the boundary of the strong coupling condition, the PL is very significantly modified.\\

\noindent\textbf{RhB outside the cavity.}
We next investigated a non-aggregated dye system, we switched from TDBC to a custom thiolated variant of a standard laser dye, RhB. (Briefly, this involved modifying the RhB molecule so as to include an alkynethiol, the original idea being to make a version of RhB capable of self-assembly.) For this RhB `outside the cavity' sample the RhB layer was deposited on top of the cavity by drop casting, see details in SI. The PL dispersion data are shown in the right-hand column of figure \ref{fig:schematics_and_PL}. The situation for this dye - as for many dyes - is somewhat complicated by the multi-oscillator nature of the excitonic transition. Nonetheless, the data show that the PL tracks the lower polariton, even though the dye is located outside the cavity.\\

For this system we find the bare cavity mode linewidth to be $K = 0.18\,\pm 0.02$ eV whilst the RhB linewidth we determined to be $\Gamma=0.21\,\pm 0.02$ eV. From our match of a coupled oscillator model to our PL and reflectance data for this system we find that the Rabi splitting is 0.20 $\pm$ 0.04 eV, so that we find $\Omega_{\textrm{R}}/((K+\Gamma)/2) = 1.0$ $\pm$ 0.4. Although the error bar is substantial, we are roughly at the strong coupling condition, consistent with what we see from our PL data (shown in the right-hand column of figure \ref{fig:schematics_and_PL}), i.e. that the PL clearly tracks the lower polariton branch. Again, as for the TDBC outside the cavity, PL from the RhB film does not pass through the cavity on its way to the detector, there is no filtering effect.\\

In the supplementary information we show, for each cavity structure, the calculated power dissipated in the dye layer when illuminated from above. Again, these data are presented in the form of dispersion plots. For the TDBC inside the cavity the dispersion takes the form of the lower polariton. For the TDBC outside the cavity and the RhB outside the cavity, there is little if any evidence that power dissipated is influenced by the lower polariton. This is perhaps not unexpected since in this case the dye layers are readily dominated by bare exciton absorption. Nonetheless, it would be interesting to extend this aspect of our investigation.

Elsewhere we have indicated that -- based on photoluminescence measurements -- `simply' meeting the strong coupling condition, \ref{eq:sc_condition} may not be enough~\cite{Menghrajani_arXiv_2211.08300}, and suggested that a second criterion might need to be applied, that the cavity finesse needs to be $\mathcal{F}>5$. As indicated in the supplementary information, all three cavities we studied here had $\mathcal{F}>10$, thus meeting this extra criterion.

\section*{Summary}
In summary, our study successfully demonstrated the attainment of strong coupling by positioning molecules outside a Fabry-Perot cavity. This approach offers all of the attributes discussed in table 1, making it interesting and valuable platform for polaritonic chemistry. 

There is of course an obvious downside to our approach, it places the molecules in a low field region of the cavity mode, thus reducing the attainable coupling strength. We have however shown that despite this problem, strong coupling can still be achieved. Nonetheless, we have done this by using materials with a combination of high concentration and high oscillator strength.m It would be useful if there were a way to boost the strength of the coupling, e.g. for vibrational coupling where the molecular resonances involved are in the infrared, and are the resonances that are so important to polaritonic chemistry. This might additionally have the benefit of enabling a single monolayer of molecules exhibit strong coupling (for an excitonic resonance). To achieve this we suggest that two molecular systems could be used, possibly even employing the same molecules. One set of molecules is placed \textit{inside} the cavity. These molecules are not involved in any external chemistry, and could be dispersed in a solid matrix. The role of these `internal' molecules is simply to boost the coupling strength. The second set of molecules is then deposited on the outside of the cavity as we have indicated. One could also consider using an hierarchical approach~\cite{Bisht_NL_2019_19_189}.

The ability to access and control strong coupling dynamics outside a conventional cavity structure broadens the scope of possibilities for polaritonic chemistry. As we have indicated above, this may be possible down to the single monolayer regime, making it attractive for studies of catalysis. We hope our findings will lead to a promising avenue for future research and to new 
applications of strong coupling in various fields.

\section*{\label{sec:level1}Acknowledgements}
The authors are grateful for and acknowledge the financial support of the Leverhulme Trust, associated with the research grant ``Synthetic Biological Control of Quantum Optics''. K.S.M. also acknowledges the support of Royal Society International Exchange grant (119893R).
We thank Adarsh Vasista, Wai Jue Tan, Philip Thomas, Marie Rider, Felipe Herrera, and William Wardley for many fruitful discussions.
The authors also acknowledge the support of European Research Council through the Photmat project 
(ERC-2016-AdG-742222
:www.photmat.eu).

\bibliography{refs_outside}

\begin{thebibliography}{25}%
\makeatletter
\providecommand \@ifxundefined [1]{%
 \@ifx{#1\undefined}
}%
\providecommand \@ifnum [1]{%
 \ifnum #1\expandafter \@firstoftwo
 \else \expandafter \@secondoftwo
 \fi
}%
\providecommand \@ifx [1]{%
 \ifx #1\expandafter \@firstoftwo
 \else \expandafter \@secondoftwo
 \fi
}%
\providecommand \natexlab [1]{#1}%
\providecommand \enquote  [1]{``#1''}%
\providecommand \bibnamefont  [1]{#1}%
\providecommand \bibfnamefont [1]{#1}%
\providecommand \citenamefont [1]{#1}%
\providecommand \href@noop [0]{\@secondoftwo}%
\providecommand \href [0]{\begingroup \@sanitize@url \@href}%
\providecommand \@href[1]{\@@startlink{#1}\@@href}%
\providecommand \@@href[1]{\endgroup#1\@@endlink}%
\providecommand \@sanitize@url [0]{\catcode `\\12\catcode `\$12\catcode `\&12\catcode `\#12\catcode `\^12\catcode `\_12\catcode `\%12\relax}%
\providecommand \@@startlink[1]{}%
\providecommand \@@endlink[0]{}%
\providecommand \url  [0]{\begingroup\@sanitize@url \@url }%
\providecommand \@url [1]{\endgroup\@href {#1}{\urlprefix }}%
\providecommand \urlprefix  [0]{URL }%
\providecommand \Eprint [0]{\href }%
\providecommand \doibase [0]{https://doi.org/}%
\providecommand \selectlanguage [0]{\@gobble}%
\providecommand \bibinfo  [0]{\@secondoftwo}%
\providecommand \bibfield  [0]{\@secondoftwo}%
\providecommand \translation [1]{[#1]}%
\providecommand \BibitemOpen [0]{}%
\providecommand \bibitemStop [0]{}%
\providecommand \bibitemNoStop [0]{.\EOS\space}%
\providecommand \EOS [0]{\spacefactor3000\relax}%
\providecommand \BibitemShut  [1]{\csname bibitem#1\endcsname}%
\let\auto@bib@innerbib\@empty
\bibitem [{\citenamefont {Ebbesen}(2016)}]{Ebbesen_ACS_Accounts_2016_49_2403}%
  \BibitemOpen
  \bibfield  {author} {\bibinfo {author} {\bibfnamefont {T.~W.}\ \bibnamefont {Ebbesen}},\ }\bibfield  {title} {\bibinfo {title} {Hybrid light--matter states in a molecular and material science perspective},\ }\href@noop {} {\bibfield  {journal} {\bibinfo  {journal} {Accounts of Chemical Research}\ }\textbf {\bibinfo {volume} {49}},\ \bibinfo {pages} {2403} (\bibinfo {year} {2016})}\BibitemShut {NoStop}%
\bibitem [{\citenamefont {Herrera}\ and\ \citenamefont {Owrutsky}(2020)}]{Herrera_JCP_2020_152_100902}%
  \BibitemOpen
  \bibfield  {author} {\bibinfo {author} {\bibfnamefont {F.}~\bibnamefont {Herrera}}\ and\ \bibinfo {author} {\bibfnamefont {J.}~\bibnamefont {Owrutsky}},\ }\bibfield  {title} {\bibinfo {title} {Molecular polaritons for controlling chemistry with quantum optics},\ }\href@noop {} {\bibfield  {journal} {\bibinfo  {journal} {The Journal of Chemical Physics}\ }\textbf {\bibinfo {volume} {152}},\ \bibinfo {pages} {100902} (\bibinfo {year} {2020})}\BibitemShut {NoStop}%
\bibitem [{\citenamefont {Simpkins}\ \emph {et~al.}(2021)\citenamefont {Simpkins}, \citenamefont {Dunkelberger},\ and\ \citenamefont {Owrutsky}}]{Simpkins_JPCC_2021_125_19081}%
  \BibitemOpen
  \bibfield  {author} {\bibinfo {author} {\bibfnamefont {B.~S.}\ \bibnamefont {Simpkins}}, \bibinfo {author} {\bibfnamefont {A.~D.}\ \bibnamefont {Dunkelberger}},\ and\ \bibinfo {author} {\bibfnamefont {J.~C.}\ \bibnamefont {Owrutsky}},\ }\bibfield  {title} {\bibinfo {title} {Mode-specific chemistry through vibrational strong coupling (or a wish come true)},\ }\href@noop {} {\bibfield  {journal} {\bibinfo  {journal} {The Journal of Physical Chemistry C}\ }\textbf {\bibinfo {volume} {125}},\ \bibinfo {pages} {19081} (\bibinfo {year} {2021})}\BibitemShut {NoStop}%
\bibitem [{\citenamefont {Ahn}\ \emph {et~al.}(2023)\citenamefont {Ahn}, \citenamefont {Triana}, \citenamefont {Recabal}, \citenamefont {Herrera},\ and\ \citenamefont {Simpkins}}]{Ahn_Science_2023_380_6650}%
  \BibitemOpen
  \bibfield  {author} {\bibinfo {author} {\bibfnamefont {W.}~\bibnamefont {Ahn}}, \bibinfo {author} {\bibfnamefont {J.~F.}\ \bibnamefont {Triana}}, \bibinfo {author} {\bibfnamefont {F.}~\bibnamefont {Recabal}}, \bibinfo {author} {\bibfnamefont {F.}~\bibnamefont {Herrera}},\ and\ \bibinfo {author} {\bibfnamefont {B.~S.}\ \bibnamefont {Simpkins}},\ }\bibfield  {title} {\bibinfo {title} {Modification of ground-state chemical reactivity via light-matter coherence in infrared cavities},\ }\href@noop {} {\bibfield  {journal} {\bibinfo  {journal} {Science}\ }\textbf {\bibinfo {volume} {380}},\ \bibinfo {pages} {1165} (\bibinfo {year} {2023})}\BibitemShut {NoStop}%
\bibitem [{\citenamefont {Baieva}\ \emph {et~al.}(2013)\citenamefont {Baieva}, \citenamefont {Ihalainen},\ and\ \citenamefont {Toppari}}]{Baieva_JCP_2013_138_044707}%
  \BibitemOpen
  \bibfield  {author} {\bibinfo {author} {\bibfnamefont {S.}~\bibnamefont {Baieva}}, \bibinfo {author} {\bibfnamefont {J.~A.}\ \bibnamefont {Ihalainen}},\ and\ \bibinfo {author} {\bibfnamefont {J.~J.}\ \bibnamefont {Toppari}},\ }\bibfield  {title} {\bibinfo {title} {Strong coupling between surface plasmon polaritons and $\beta$-carotene in nanolayered system},\ }\href@noop {} {\bibfield  {journal} {\bibinfo  {journal} {The Journal of Chemical Physics}\ }\textbf {\bibinfo {volume} {138}},\ \bibinfo {pages} {044707} (\bibinfo {year} {2013})}\BibitemShut {NoStop}%
\bibitem [{\citenamefont {T{\"o}rm{\"a}}\ and\ \citenamefont {Barnes}(2015)}]{Torma_RepProgPhys_2015_78_013901}%
  \BibitemOpen
  \bibfield  {author} {\bibinfo {author} {\bibfnamefont {P.}~\bibnamefont {T{\"o}rm{\"a}}}\ and\ \bibinfo {author} {\bibfnamefont {W.~L.}\ \bibnamefont {Barnes}},\ }\bibfield  {title} {\bibinfo {title} {Strong coupling between surface plasmon polaritons and emitters: a review},\ }\href@noop {} {\bibfield  {journal} {\bibinfo  {journal} {Reports on Progress in Physics}\ }\textbf {\bibinfo {volume} {78}},\ \bibinfo {pages} {013901} (\bibinfo {year} {2015})}\BibitemShut {NoStop}%
\bibitem [{\citenamefont {Vasista}\ and\ \citenamefont {Barnes}(2020)}]{Vasista_NL_2020_20_1766}%
  \BibitemOpen
  \bibfield  {author} {\bibinfo {author} {\bibfnamefont {A.}~\bibnamefont {Vasista}}\ and\ \bibinfo {author} {\bibfnamefont {W.~L.}\ \bibnamefont {Barnes}},\ }\bibfield  {title} {\bibinfo {title} {Molecular monolayer strong coupling in dielectric soft microcavities},\ }\href@noop {} {\bibfield  {journal} {\bibinfo  {journal} {Nano Letters}\ }\textbf {\bibinfo {volume} {20}},\ \bibinfo {pages} {1766} (\bibinfo {year} {2020})}\BibitemShut {NoStop}%
\bibitem [{\citenamefont {Yadav}\ \emph {et~al.}(2020)\citenamefont {Yadav}, \citenamefont {Otten}, \citenamefont {Wang}, \citenamefont {Cortes}, \citenamefont {Gosztola}, \citenamefont {Wiederrecht}, \citenamefont {Gray}, \citenamefont {Odom},\ and\ \citenamefont {Basu}}]{Yadav_NL_2020_20_5043}%
  \BibitemOpen
  \bibfield  {author} {\bibinfo {author} {\bibfnamefont {R.~K.}\ \bibnamefont {Yadav}}, \bibinfo {author} {\bibfnamefont {M.}~\bibnamefont {Otten}}, \bibinfo {author} {\bibfnamefont {W.}~\bibnamefont {Wang}}, \bibinfo {author} {\bibfnamefont {C.~L.}\ \bibnamefont {Cortes}}, \bibinfo {author} {\bibfnamefont {D.~J.}\ \bibnamefont {Gosztola}}, \bibinfo {author} {\bibfnamefont {G.~P.}\ \bibnamefont {Wiederrecht}}, \bibinfo {author} {\bibfnamefont {S.~K.}\ \bibnamefont {Gray}}, \bibinfo {author} {\bibfnamefont {T.~W.}\ \bibnamefont {Odom}},\ and\ \bibinfo {author} {\bibfnamefont {J.~K.}\ \bibnamefont {Basu}},\ }\bibfield  {title} {\bibinfo {title} {Strongly coupled exciton--surface lattice resonances engineer long-range energy propagation},\ }\href@noop {} {\bibfield  {journal} {\bibinfo  {journal} {Nano Letters}\ }\textbf {\bibinfo {volume} {20}},\ \bibinfo {pages} {5043} (\bibinfo {year} {2020})}\BibitemShut {NoStop}%
\bibitem [{\citenamefont {Verdelli}\ \emph {et~al.}(2022)\citenamefont {Verdelli}, \citenamefont {Schulpen}, \citenamefont {Baldi},\ and\ \citenamefont {Rivas}}]{Verdelli_JPCC_2022_126_7143}%
  \BibitemOpen
  \bibfield  {author} {\bibinfo {author} {\bibfnamefont {F.}~\bibnamefont {Verdelli}}, \bibinfo {author} {\bibfnamefont {J.~J. P.~M.}\ \bibnamefont {Schulpen}}, \bibinfo {author} {\bibfnamefont {A.}~\bibnamefont {Baldi}},\ and\ \bibinfo {author} {\bibfnamefont {J.~G.}\ \bibnamefont {Rivas}},\ }\bibfield  {title} {\bibinfo {title} {Chasing vibro-polariton fingerprints in infrared and raman spectra using surface lattice resonances on extended metasurfaces},\ }\href@noop {} {\bibfield  {journal} {\bibinfo  {journal} {The Journal of Physical Chemistry C}\ }\textbf {\bibinfo {volume} {126}},\ \bibinfo {pages} {7143} (\bibinfo {year} {2022})}\BibitemShut {NoStop}%
\bibitem [{\citenamefont {Georgiou}\ \emph {et~al.}(2020)\citenamefont {Georgiou}, \citenamefont {Jayaprakash},\ and\ \citenamefont {Lidzey}}]{Georgiou_JPCL_2020_11_9893}%
  \BibitemOpen
  \bibfield  {author} {\bibinfo {author} {\bibfnamefont {K.}~\bibnamefont {Georgiou}}, \bibinfo {author} {\bibfnamefont {R.}~\bibnamefont {Jayaprakash}},\ and\ \bibinfo {author} {\bibfnamefont {D.~G.}\ \bibnamefont {Lidzey}},\ }\bibfield  {title} {\bibinfo {title} {Strong coupling of organic dyes located at the surface of a dielectric slab microcavity},\ }\href@noop {} {\bibfield  {journal} {\bibinfo  {journal} {The Journal of Physical Chemistry Letters}\ }\textbf {\bibinfo {volume} {11}},\ \bibinfo {pages} {9893} (\bibinfo {year} {2020})}\BibitemShut {NoStop}%
\bibitem [{\citenamefont {Thomas}\ \emph {et~al.}(2021)\citenamefont {Thomas}, \citenamefont {Menghrajani},\ and\ \citenamefont {Barnes}}]{Thomas_JPCL_2021_12_6914}%
  \BibitemOpen
  \bibfield  {author} {\bibinfo {author} {\bibfnamefont {P.~A.}\ \bibnamefont {Thomas}}, \bibinfo {author} {\bibfnamefont {K.~S.}\ \bibnamefont {Menghrajani}},\ and\ \bibinfo {author} {\bibfnamefont {W.~L.}\ \bibnamefont {Barnes}},\ }\bibfield  {title} {\bibinfo {title} {Cavity-free ultrastrong light-matter coupling},\ }\href@noop {} {\bibfield  {journal} {\bibinfo  {journal} {The Journal of Physical Chemistry Letters}\ }\textbf {\bibinfo {volume} {12}},\ \bibinfo {pages} {6914} (\bibinfo {year} {2021})}\BibitemShut {NoStop}%
\bibitem [{\citenamefont {Canales}\ \emph {et~al.}(2021)\citenamefont {Canales}, \citenamefont {Baranov}, \citenamefont {Antosiewicz},\ and\ \citenamefont {Shegai}}]{Canales_JCP_2021_154_024701}%
  \BibitemOpen
  \bibfield  {author} {\bibinfo {author} {\bibfnamefont {A.}~\bibnamefont {Canales}}, \bibinfo {author} {\bibfnamefont {D.~G.}\ \bibnamefont {Baranov}}, \bibinfo {author} {\bibfnamefont {T.~J.}\ \bibnamefont {Antosiewicz}},\ and\ \bibinfo {author} {\bibfnamefont {T.}~\bibnamefont {Shegai}},\ }\bibfield  {title} {\bibinfo {title} {Abundance of cavity-free polaritonic states in resonant materials and nanostructures},\ }\href@noop {} {\bibfield  {journal} {\bibinfo  {journal} {The Journal of Chemical Physics}\ }\textbf {\bibinfo {volume} {154}},\ \bibinfo {pages} {024701} (\bibinfo {year} {2021})}\BibitemShut {NoStop}%
\bibitem [{\citenamefont {Ahn}\ \emph {et~al.}(2010)\citenamefont {Ahn}, \citenamefont {Kimerling},\ and\ \citenamefont {Michel}}]{Ahn_JLT_2010_28_3387}%
  \BibitemOpen
  \bibfield  {author} {\bibinfo {author} {\bibfnamefont {D.}~\bibnamefont {Ahn}}, \bibinfo {author} {\bibfnamefont {L.~C.}\ \bibnamefont {Kimerling}},\ and\ \bibinfo {author} {\bibfnamefont {J.}~\bibnamefont {Michel}},\ }\bibfield  {title} {\bibinfo {title} {Evanescent coupling device design for waveguide-integrated group iv photodetectors},\ }\href@noop {} {\bibfield  {journal} {\bibinfo  {journal} {J. Lightwave Technol.}\ }\textbf {\bibinfo {volume} {28}},\ \bibinfo {pages} {3387} (\bibinfo {year} {2010})}\BibitemShut {NoStop}%
\bibitem [{\citenamefont {Moon}\ \emph {et~al.}(2004)\citenamefont {Moon}, \citenamefont {Park}, \citenamefont {Lee}, \citenamefont {An},\ and\ \citenamefont {Lee}}]{Moon_APL_2004_84_4547}%
  \BibitemOpen
  \bibfield  {author} {\bibinfo {author} {\bibfnamefont {H.-J.}\ \bibnamefont {Moon}}, \bibinfo {author} {\bibfnamefont {G.-W.}\ \bibnamefont {Park}}, \bibinfo {author} {\bibfnamefont {S.-B.}\ \bibnamefont {Lee}}, \bibinfo {author} {\bibfnamefont {K.}~\bibnamefont {An}},\ and\ \bibinfo {author} {\bibfnamefont {J.-H.}\ \bibnamefont {Lee}},\ }\bibfield  {title} {\bibinfo {title} {{Waveguide mode lasing via evanescent-wave-coupled gain from a thin cylindrical shell resonator}},\ }\href@noop {} {\bibfield  {journal} {\bibinfo  {journal} {Applied Physics Letters}\ }\textbf {\bibinfo {volume} {84}},\ \bibinfo {pages} {4547} (\bibinfo {year} {2004})}\BibitemShut {NoStop}%
\bibitem [{\citenamefont {Fang}\ \emph {et~al.}(2007)\citenamefont {Fang}, \citenamefont {Park}, \citenamefont {hao Kuo}, \citenamefont {Jones}, \citenamefont {Cohen}, \citenamefont {Liang}, \citenamefont {Raday}, \citenamefont {Sysak}, \citenamefont {Paniccia},\ and\ \citenamefont {Bowers}}]{Fang_MatToday_2007_10_28}%
  \BibitemOpen
  \bibfield  {author} {\bibinfo {author} {\bibfnamefont {A.~W.}\ \bibnamefont {Fang}}, \bibinfo {author} {\bibfnamefont {H.}~\bibnamefont {Park}}, \bibinfo {author} {\bibfnamefont {Y.}~\bibnamefont {hao Kuo}}, \bibinfo {author} {\bibfnamefont {R.}~\bibnamefont {Jones}}, \bibinfo {author} {\bibfnamefont {O.}~\bibnamefont {Cohen}}, \bibinfo {author} {\bibfnamefont {D.}~\bibnamefont {Liang}}, \bibinfo {author} {\bibfnamefont {O.}~\bibnamefont {Raday}}, \bibinfo {author} {\bibfnamefont {M.~N.}\ \bibnamefont {Sysak}}, \bibinfo {author} {\bibfnamefont {M.~J.}\ \bibnamefont {Paniccia}},\ and\ \bibinfo {author} {\bibfnamefont {J.~E.}\ \bibnamefont {Bowers}},\ }\bibfield  {title} {\bibinfo {title} {Hybrid silicon evanescent devices},\ }\href@noop {} {\bibfield  {journal} {\bibinfo  {journal} {Materials Today}\ }\textbf {\bibinfo {volume} {10}},\ \bibinfo {pages} {28} (\bibinfo {year} {2007})}\BibitemShut {NoStop}%
\bibitem [{\citenamefont {Balasubrahmaniyam}\ \emph {et~al.}(2023)\citenamefont {Balasubrahmaniyam}, \citenamefont {Simkhovich}, \citenamefont {Golombek}, \citenamefont {Sandik}, \citenamefont {Ankonina},\ and\ \citenamefont {Schwartz}}]{Balasubrahmaniyam_NatMat_2023_22_338}%
  \BibitemOpen
  \bibfield  {author} {\bibinfo {author} {\bibfnamefont {M.}~\bibnamefont {Balasubrahmaniyam}}, \bibinfo {author} {\bibfnamefont {A.}~\bibnamefont {Simkhovich}}, \bibinfo {author} {\bibfnamefont {A.}~\bibnamefont {Golombek}}, \bibinfo {author} {\bibfnamefont {G.}~\bibnamefont {Sandik}}, \bibinfo {author} {\bibfnamefont {G.}~\bibnamefont {Ankonina}},\ and\ \bibinfo {author} {\bibfnamefont {T.}~\bibnamefont {Schwartz}},\ }\bibfield  {title} {\bibinfo {title} {From enhanced diffusion to ultrafast ballistic motion of hybrid light--matter excitations},\ }\href@noop {} {\bibfield  {journal} {\bibinfo  {journal} {Nature Materials}\ }\textbf {\bibinfo {volume} {22}},\ \bibinfo {pages} {338} (\bibinfo {year} {2023})}\BibitemShut {NoStop}%
\bibitem [{\citenamefont {Dintinger}\ \emph {et~al.}(2005)\citenamefont {Dintinger}, \citenamefont {Klein}, \citenamefont {Bustos}, \citenamefont {Barnes},\ and\ \citenamefont {Ebbesen}}]{Dintinger_PRB_2005_71_035424}%
  \BibitemOpen
  \bibfield  {author} {\bibinfo {author} {\bibfnamefont {J.}~\bibnamefont {Dintinger}}, \bibinfo {author} {\bibfnamefont {S.}~\bibnamefont {Klein}}, \bibinfo {author} {\bibfnamefont {F.}~\bibnamefont {Bustos}}, \bibinfo {author} {\bibfnamefont {W.~L.}\ \bibnamefont {Barnes}},\ and\ \bibinfo {author} {\bibfnamefont {T.~W.}\ \bibnamefont {Ebbesen}},\ }\bibfield  {title} {\bibinfo {title} {Strong coupling between surface plasmon-polaritons and organic molecules in subwavelength hole arrays},\ }\href@noop {} {\bibfield  {journal} {\bibinfo  {journal} {Physical Review B}\ }\textbf {\bibinfo {volume} {71}},\ \bibinfo {pages} {035424} (\bibinfo {year} {2005})}\BibitemShut {NoStop}%
\bibitem [{\citenamefont {Hobson}\ \emph {et~al.}(2002)\citenamefont {Hobson}, \citenamefont {Barnes}, \citenamefont {Lidzey}, \citenamefont {Gehring}, \citenamefont {Whittaker}, \citenamefont {Skolnick},\ and\ \citenamefont {Walker}}]{Hobson_APL_2002_81_3519}%
  \BibitemOpen
  \bibfield  {author} {\bibinfo {author} {\bibfnamefont {P.~A.}\ \bibnamefont {Hobson}}, \bibinfo {author} {\bibfnamefont {W.~L.}\ \bibnamefont {Barnes}}, \bibinfo {author} {\bibfnamefont {D.~G.}\ \bibnamefont {Lidzey}}, \bibinfo {author} {\bibfnamefont {G.~A.}\ \bibnamefont {Gehring}}, \bibinfo {author} {\bibfnamefont {D.~M.}\ \bibnamefont {Whittaker}}, \bibinfo {author} {\bibfnamefont {M.~S.}\ \bibnamefont {Skolnick}},\ and\ \bibinfo {author} {\bibfnamefont {S.}~\bibnamefont {Walker}},\ }\bibfield  {title} {\bibinfo {title} {Strong exciton--photon coupling in a low-q all-metal mirror microcavity},\ }\href@noop {} {\bibfield  {journal} {\bibinfo  {journal} {Applied Physics Letters}\ }\textbf {\bibinfo {volume} {81}},\ \bibinfo {pages} {3519} (\bibinfo {year} {2002})}\BibitemShut {NoStop}%
\bibitem [{\citenamefont {Wers{\"a}ll}\ \emph {et~al.}(2019)\citenamefont {Wers{\"a}ll}, \citenamefont {Munkhbat}, \citenamefont {Baranov}, \citenamefont {Herrera}, \citenamefont {Cao}, \citenamefont {Antosiewicz},\ and\ \citenamefont {Shegai}}]{Wersall_ACSPhot_2019_6_2570}%
  \BibitemOpen
  \bibfield  {author} {\bibinfo {author} {\bibfnamefont {M.}~\bibnamefont {Wers{\"a}ll}}, \bibinfo {author} {\bibfnamefont {B.}~\bibnamefont {Munkhbat}}, \bibinfo {author} {\bibfnamefont {D.~G.}\ \bibnamefont {Baranov}}, \bibinfo {author} {\bibfnamefont {F.}~\bibnamefont {Herrera}}, \bibinfo {author} {\bibfnamefont {J.}~\bibnamefont {Cao}}, \bibinfo {author} {\bibfnamefont {T.~J.}\ \bibnamefont {Antosiewicz}},\ and\ \bibinfo {author} {\bibfnamefont {T.}~\bibnamefont {Shegai}},\ }\bibfield  {title} {\bibinfo {title} {Correlative dark-field and photoluminescence spectroscopy of individual plasmon--molecule hybrid nanostructures in a strong coupling regime},\ }\href@noop {} {\bibfield  {journal} {\bibinfo  {journal} {ACS Photonics}\ }\textbf {\bibinfo {volume} {6}},\ \bibinfo {pages} {2570} (\bibinfo {year} {2019})}\BibitemShut {NoStop}%
\bibitem [{\citenamefont {Vasista}\ \emph {et~al.}(2021)\citenamefont {Vasista}, \citenamefont {Menghrajani},\ and\ \citenamefont {Barnes}}]{Vasista_Nanoscale_2021_13_14497}%
  \BibitemOpen
  \bibfield  {author} {\bibinfo {author} {\bibfnamefont {A.~B.}\ \bibnamefont {Vasista}}, \bibinfo {author} {\bibfnamefont {K.~S.}\ \bibnamefont {Menghrajani}},\ and\ \bibinfo {author} {\bibfnamefont {W.~L.}\ \bibnamefont {Barnes}},\ }\bibfield  {title} {\bibinfo {title} {Polariton assisted photoemission from a layered molecular material: role of vibrational states and molecular absorption},\ }\href@noop {} {\bibfield  {journal} {\bibinfo  {journal} {Nanoscale}\ }\textbf {\bibinfo {volume} {13}},\ \bibinfo {pages} {14497} (\bibinfo {year} {2021})}\BibitemShut {NoStop}%
\bibitem [{\citenamefont {Rider}\ and\ \citenamefont {Barnes}(2021)}]{Rider_CP_2021_62_217}%
  \BibitemOpen
  \bibfield  {author} {\bibinfo {author} {\bibfnamefont {M.~S.}\ \bibnamefont {Rider}}\ and\ \bibinfo {author} {\bibfnamefont {W.~L.}\ \bibnamefont {Barnes}},\ }\bibfield  {title} {\bibinfo {title} {Something from nothing: linking molecules with virtual light},\ }\href@noop {} {\bibfield  {journal} {\bibinfo  {journal} {Contemporary Physics}\ }\textbf {\bibinfo {volume} {62}},\ \bibinfo {pages} {217} (\bibinfo {year} {2021})}\BibitemShut {NoStop}%
\bibitem [{\citenamefont {Heilmann}\ \emph {et~al.}(2020)\citenamefont {Heilmann}, \citenamefont {Vakevainen}, \citenamefont {Martikainen},\ and\ \citenamefont {Torma}}]{Heilmann_Nanophot_2020_9_267}%
  \BibitemOpen
  \bibfield  {author} {\bibinfo {author} {\bibfnamefont {R.}~\bibnamefont {Heilmann}}, \bibinfo {author} {\bibfnamefont {A.~I.}\ \bibnamefont {Vakevainen}}, \bibinfo {author} {\bibfnamefont {J.-P.}\ \bibnamefont {Martikainen}},\ and\ \bibinfo {author} {\bibfnamefont {P.}~\bibnamefont {Torma}},\ }\bibfield  {title} {\bibinfo {title} {Strong coupling between organic dye molecules and lattice modes of a dielectric nanoparticle array},\ }\href@noop {} {\bibfield  {journal} {\bibinfo  {journal} {Nanophotonics}\ }\textbf {\bibinfo {volume} {9}},\ \bibinfo {pages} {267} (\bibinfo {year} {2020})}\BibitemShut {NoStop}%
\bibitem [{\citenamefont {Menghrajani}\ \emph {et~al.}(2022)\citenamefont {Menghrajani}, \citenamefont {Vasista}, \citenamefont {Tan},\ and\ \citenamefont {Barnes}}]{Menghrajani_arXiv_2211.08300}%
  \BibitemOpen
  \bibfield  {author} {\bibinfo {author} {\bibfnamefont {K.~S.}\ \bibnamefont {Menghrajani}}, \bibinfo {author} {\bibfnamefont {A.~B.}\ \bibnamefont {Vasista}}, \bibinfo {author} {\bibfnamefont {W.~J.}\ \bibnamefont {Tan}},\ and\ \bibinfo {author} {\bibfnamefont {W.~L.}\ \bibnamefont {Barnes}},\ }\href@noop {} {\bibinfo {title} {Molecular strong coupling: an exploration employing open, half- and full planar cavities, arxiv 2211.08300}} (\bibinfo {year} {2022})\BibitemShut {NoStop}%
\bibitem [{\citenamefont {Bisht}\ \emph {et~al.}(2018)\citenamefont {Bisht}, \citenamefont {Cuadra}, \citenamefont {Wers{\"a}ll}, \citenamefont {Canales}, \citenamefont {Antosiewicz},\ and\ \citenamefont {Shegai}}]{Bisht_NL_2019_19_189}%
  \BibitemOpen
  \bibfield  {author} {\bibinfo {author} {\bibfnamefont {A.}~\bibnamefont {Bisht}}, \bibinfo {author} {\bibfnamefont {J.}~\bibnamefont {Cuadra}}, \bibinfo {author} {\bibfnamefont {M.}~\bibnamefont {Wers{\"a}ll}}, \bibinfo {author} {\bibfnamefont {A.}~\bibnamefont {Canales}}, \bibinfo {author} {\bibfnamefont {T.~J.}\ \bibnamefont {Antosiewicz}},\ and\ \bibinfo {author} {\bibfnamefont {T.}~\bibnamefont {Shegai}},\ }\bibfield  {title} {\bibinfo {title} {Collective strong light-matter coupling in hierarchical microcavity-plasmon-exciton systems},\ }\href@noop {} {\bibfield  {journal} {\bibinfo  {journal} {Nano Letters}\ }\textbf {\bibinfo {volume} {19}},\ \bibinfo {pages} {189} (\bibinfo {year} {2018})}\BibitemShut {NoStop}%
\bibitem [{\citenamefont {Olmon}\ \emph {et~al.}(2012)\citenamefont {Olmon}, \citenamefont {Slovick}, \citenamefont {Johnson}, \citenamefont {Shelton}, \citenamefont {Oh}, \citenamefont {Boreman},\ and\ \citenamefont {Raschke}}]{Olmon_PRB_2012_86_235147}%
  \BibitemOpen
  \bibfield  {author} {\bibinfo {author} {\bibfnamefont {R.~L.}\ \bibnamefont {Olmon}}, \bibinfo {author} {\bibfnamefont {B.}~\bibnamefont {Slovick}}, \bibinfo {author} {\bibfnamefont {T.~W.}\ \bibnamefont {Johnson}}, \bibinfo {author} {\bibfnamefont {D.}~\bibnamefont {Shelton}}, \bibinfo {author} {\bibfnamefont {S.-H.}\ \bibnamefont {Oh}}, \bibinfo {author} {\bibfnamefont {G.~D.}\ \bibnamefont {Boreman}},\ and\ \bibinfo {author} {\bibfnamefont {M.~B.}\ \bibnamefont {Raschke}},\ }\bibfield  {title} {\bibinfo {title} {Optical dielectric function of gold},\ }\href@noop {} {\bibfield  {journal} {\bibinfo  {journal} {Physical Review B}\ }\textbf {\bibinfo {volume} {86}},\ \bibinfo {pages} {235147} (\bibinfo {year} {2012})}\BibitemShut {NoStop}%
\end{thebibliography}%

\clearpage

\onecolumngrid
\vspace{1 cm}
\section*{Supplementary Information}

\section*{S1. Optical Fourier set-up}
\renewcommand{\thefigure}{S1}
Figure S1 provides a schematic of the experimental setup we employed, one that we used to acquire angle-resolved reflectance data. A white light source was focused onto the sample with a 0.8 NA 100x objective lens, the reflected signal was then collected via the same lens. Lenses L4 and L5 were used in combination to project the back-focal plane (of the objective lens) onto either the spectrometer or the camera, as determined by the position of the flip mirror FM2. Lens L6 (a flip-lens) was used to project the real plane onto the spectrometer or camera.\\

For photoluminescence (PL) measurements a beam-expanded 532 nm diode-pumped source was focused onto the sample with the objective lens, the emitted PL was collected in the back-scattering configuration. To remove the laser line, spectral edge filters were placed after lens L5. For excitation with near-normal incidence (k$\sim$0), a lens (L7) was inserted into the input path, to ensure that the laser beam was focused onto the back aperture of the objective lens. This arrangement produced an approximately parallel beam with k$\sim$0 at the sample plane.

\begin{figure*}[h!]
\centering
\includegraphics[width=0.7\linewidth]{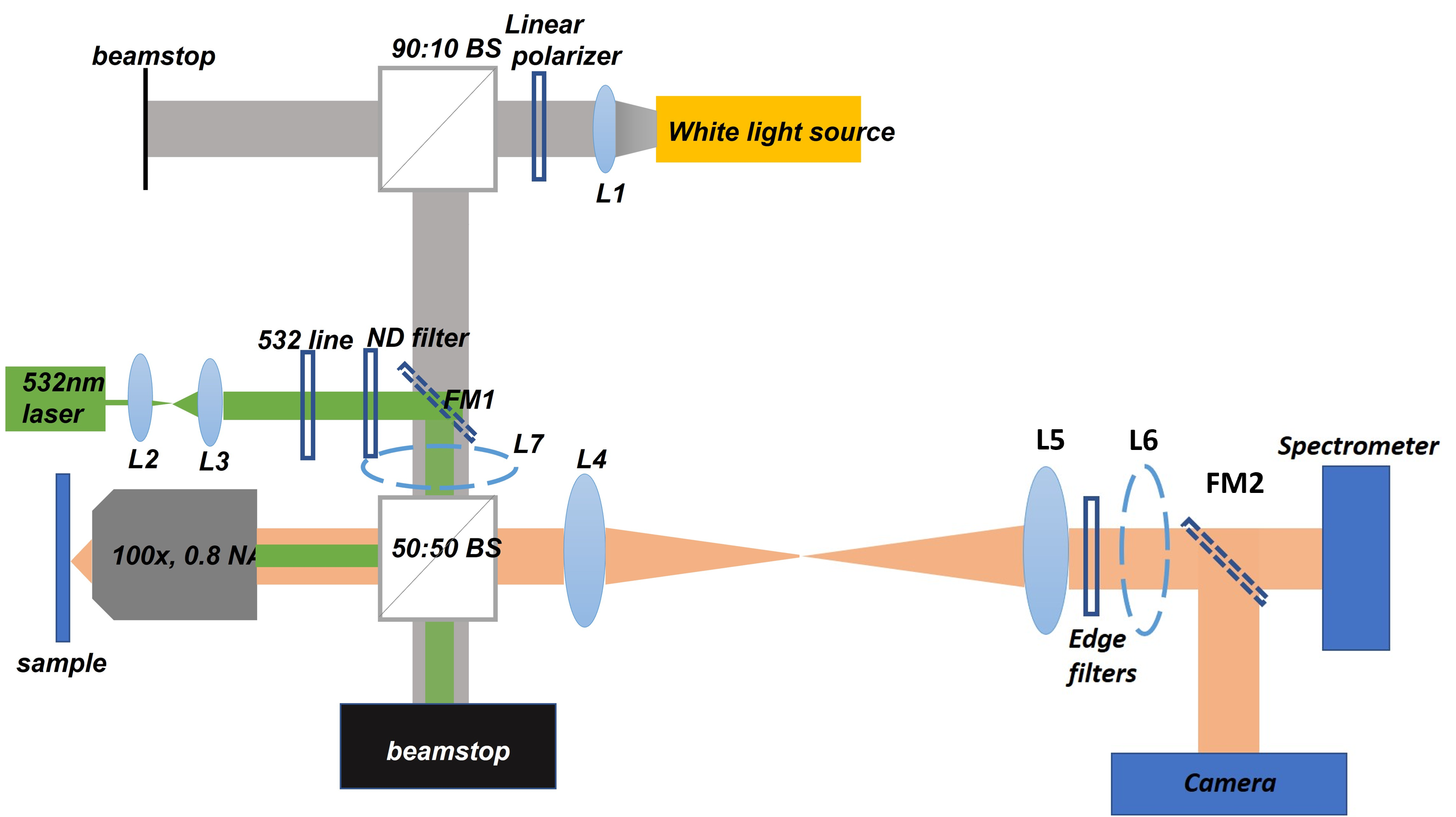}
\caption{\textbf{Schematic of the Fourier optical setup.} The function of the different elements, e.g. L5 etc, is provided in the associated text.}
\label{fig:SI_Fourier_setup}
\end{figure*}

\clearpage

\section*{S2. Photoluminescence spectra of thin films}
\renewcommand{\thefigure}{S2}

\begin{figure}[h]
\includegraphics[width=0.8\linewidth]{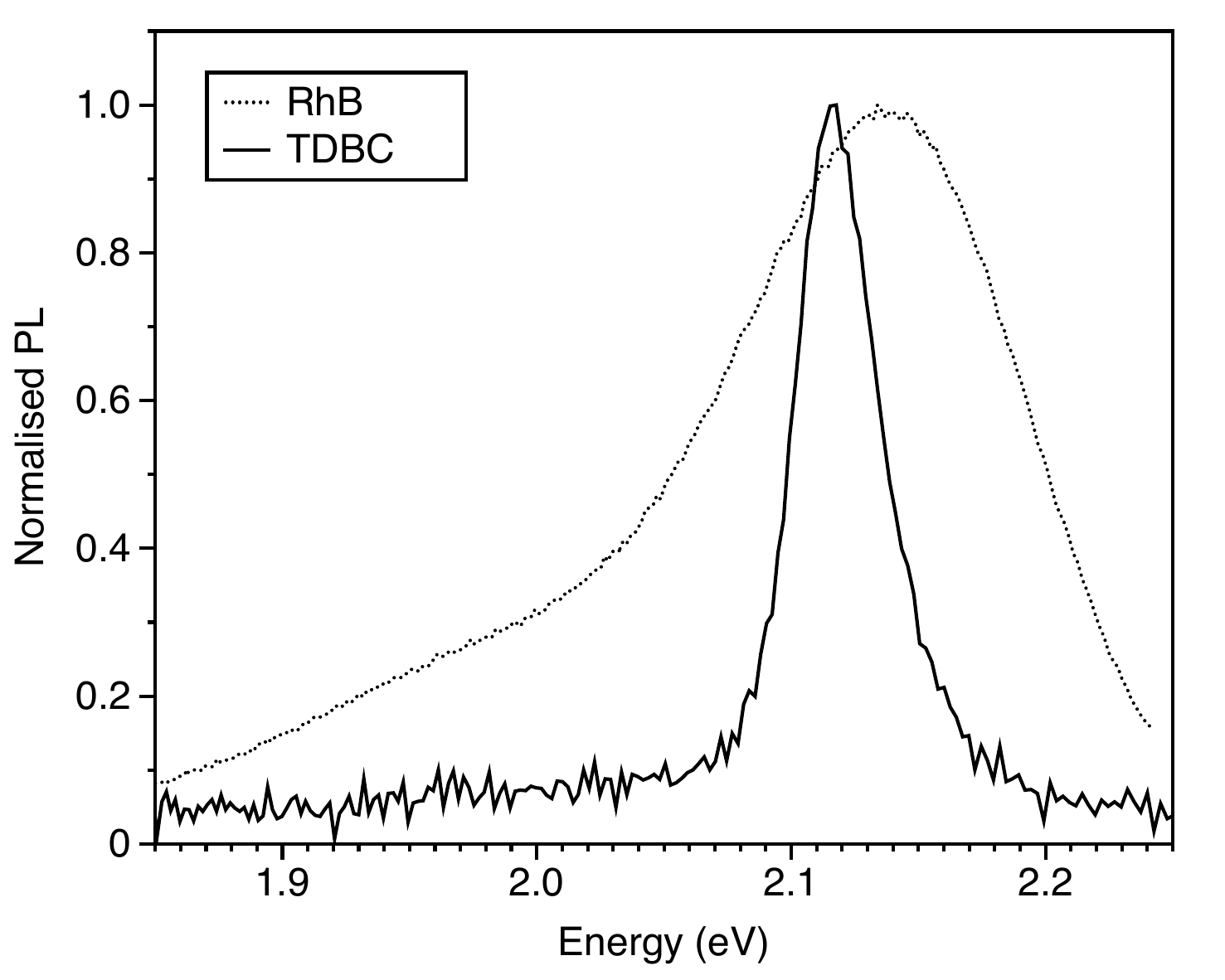}
\centering
\caption{\textbf{Reference PL spectra}. Shown as a solid line is the reference PL spectrum of a thin film of TDBC. A similar spectrum, but for a RhB film, is shown with the dotted line.}
\label{fig:TDBC_inside}
\end{figure}

\clearpage

\section*{S3. TDBC inside the cavity data}
\renewcommand{\thefigure}{S3}

\begin{figure}[b]
\includegraphics[width=0.8\linewidth]{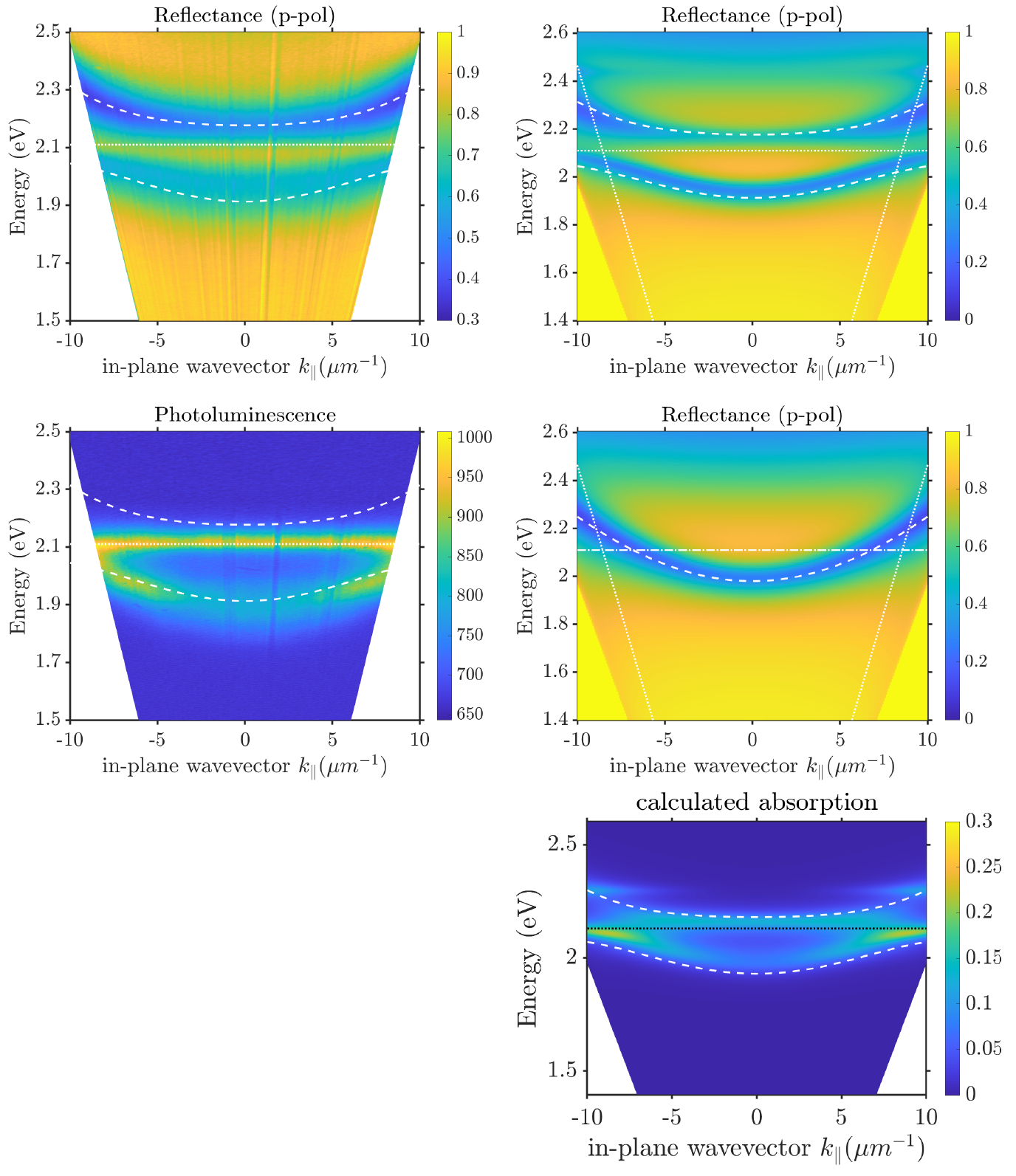}
\centering
\caption{\textbf{TDBC inside the cavity}. \textbf{Top left:} shows the measured reflectance in the form of a dispersion diagram. \textbf{Top right:} shows the calculated reflectance (transfer matrix), whilst the middle right panel shows the same calculation except that the oscillator strength has been set to zero. For completeness the PL dispersion data from the main manuscript are also shown (lower left panel. For all panels the results from a coupled oscillator model are shown as dashed white lines.}
\label{fig:TDBC_inside}
\end{figure}

The cavity here was based on a Si substrate, upon which a lower mirror of gold (approx. 40nm) was deposited. The cavity spacer was formed of two layers of PMMA, lower layer 100 nm, upper layer 247 nm, between which the 4 monolayers (8 nm) were placed. The top mirror was again gold (approx. 40nm thick). For the coupled oscillator model for this cavity the detuning was -0.15 eV, the value of $\beta$ was set to 0.65 (see ~\cite{Menghrajani_arXiv_2211.08300}), and the Rabi splitting set to $\Omega_{\textrm{R}}=0.15$ eV. The linewidth of the cavity mode was estimated from the lower right panel as $\sim$ 0.11 eV. For the calculations (right-hand panels) a Fresnel-based transfer matrix type of approach was used. The power dissipated in the dye layer when illuminated by light incident from above is shown in the lower right panel.

\clearpage

\section*{S4. TDBC outside the cavity data}
\renewcommand{\thefigure}{S4}

\begin{figure}[b]
\includegraphics[width=0.8\linewidth]{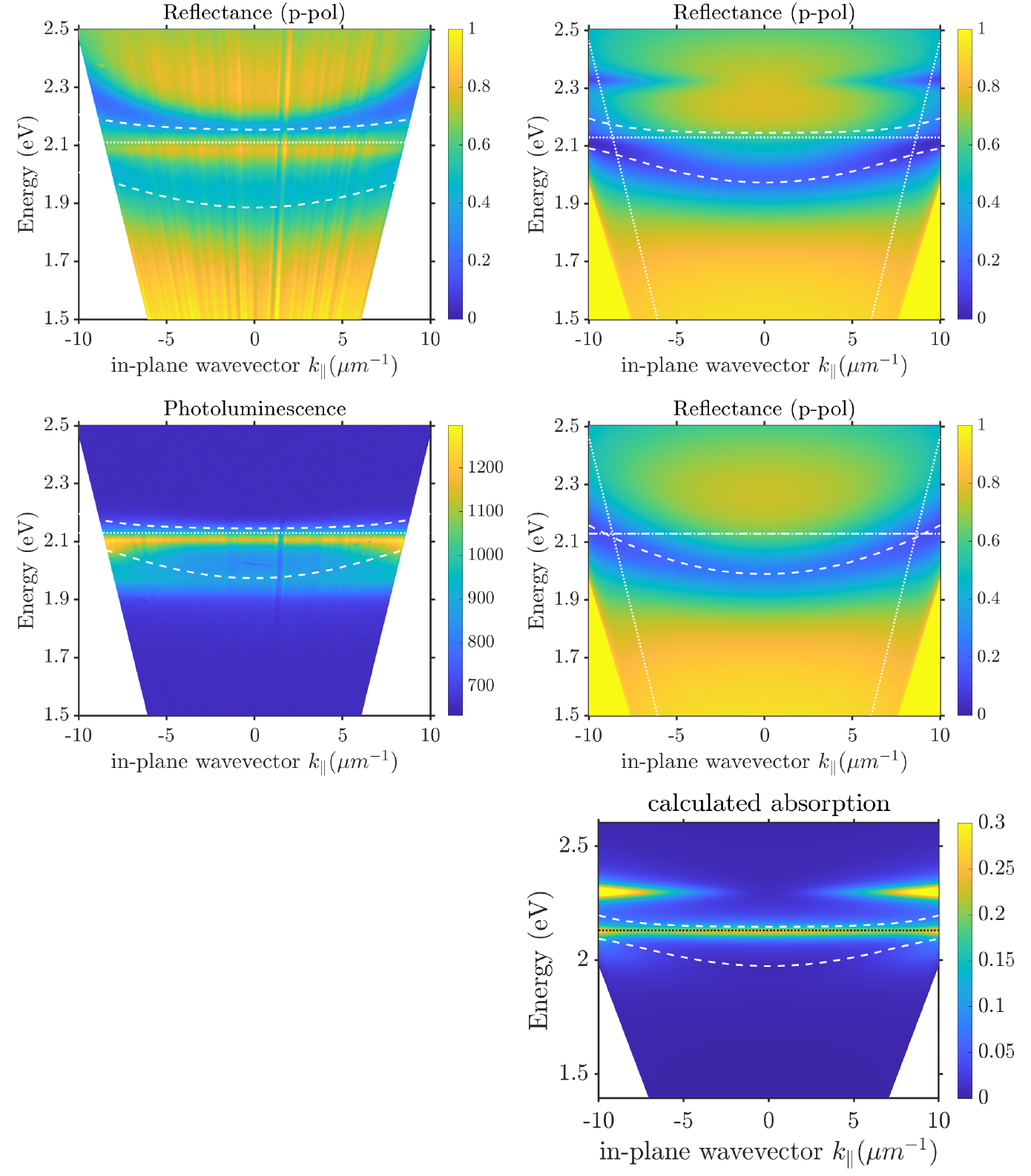}
\centering
\caption{\textbf{TDBC outside the cavity}. \textbf{Top left:} shows the measured reflectance in the form of a dispersion diagram. \textbf{Top right:} shows the calculated reflectance (transfer matrix), whilst the middle right panel shows the same calculation except that the oscillator strength has been set to zero. For completeness the PL dispersion data from the main manuscript are also shown (lower left panel. For all panels the results from a coupled oscillator model are shown as dashed white lines.}
\label{fig:TDBC_outside}
\end{figure}

The cavity here was similar to that for the TDBC inside the cavity except that (i) the spacer was just one layer of PMMA 145 nm thick, and the TDBC layer (8 nm) was placed on top of the top gold mirror. For the coupled oscillator model for this cavity the detuning was -0.14 eV, the value of $\beta$ was set to 0.4, and the Rabi splitting set to $\Omega_{\textrm{R}}=0.10$ eV. The linewidth of the cavity mode was estimated from the lower right panel as $\sim$ 0.18 eV. For the calculations (right-hand panels) a Fresnel-based transfer matrix type of approach was used. The power dissipated in the dye layer when illuminated by light incident from above is shown in the lower right panel.

\clearpage

\section*{S5. RhB outside the cavity data}
\renewcommand{\thefigure}{S5}

\begin{figure}[b]
\includegraphics[width=0.8\linewidth]{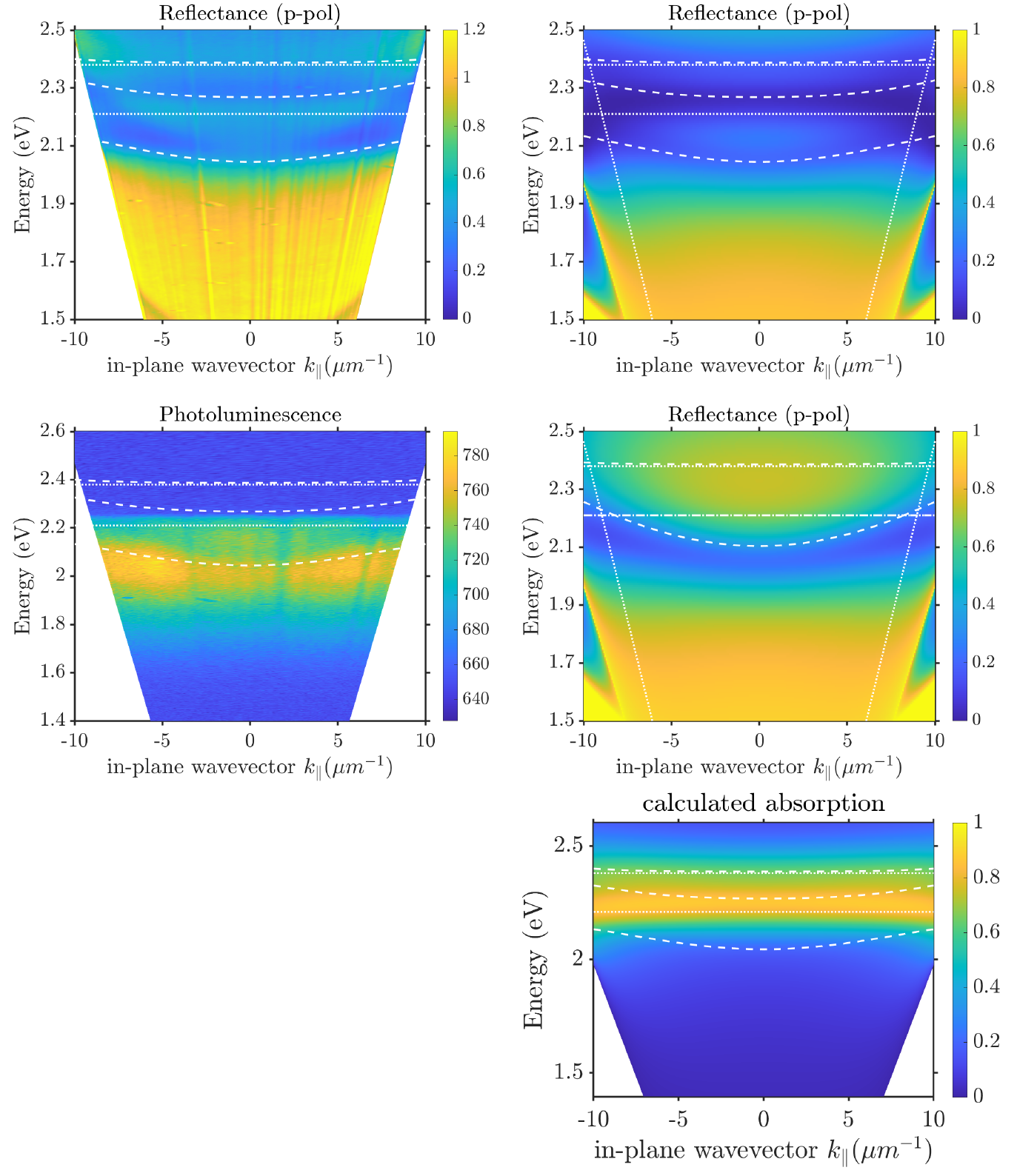}
\centering
\caption{\textbf{RhB outside the cavity}. \textbf{Top left:} shows the measured reflectance in the form of a dispersion diagram. \textbf{Top right:} shows the calculated reflectance (transfer matrix), whilst the middle right panel shows the same calculation except that the oscillator strength has been set to zero. For completeness the PL dispersion data from the main manuscript are also shown (lower left panel. For all panels the results from a coupled oscillator model are shown as dashed white lines.}
\label{fig:RhB_outside}
\end{figure}

The cavity here was very similar to that for the TDBC outside, save for the fact that RhB (150 nm) replaced the TDBC, and the PMMA thickness was 150 nm. For the coupled oscillator model for this cavity the detuning was -0.10 eV, the value of $\beta$ was set to 0.4 (see Methods), and the Rabi splitting set to $\Omega_{\textrm{R}}=0.20$ eV. Two molecular resonances were included, see below. The linewidth of the cavity mode was estimated from the lower right panel as $\sim$ 0.18 eV. For the calculations (right-hand panels) a Fresnel-based transfer matrix type of approach was used. The power dissipated in the dye layer when illuminated by light incident from above is shown in the lower right panel.

\clearpage

\section*{S6. Spectral parameters, material properties and coupled pscillator model}
\renewcommand{\thefigure}{S6}

\begin{table}[htb!]
\centering
\begin{tabular}{|p{4.0cm}|p{1.7cm}|p{1.7cm}| p{1.8cm} |p{1.6cm}|}
\hline
\hline
Cavity&$K\,$(eV)\newline{(mode-width)}& $\Gamma\,$(eV)\newline{(molecular-width)}&$\Omega_R\,$(eV)\newline{(Rabi splitting)}&$\mathcal{F}$\newline{{(Finesse)}}\\
\hline
\hline
TDBC inside &0.11$\pm0.01$&0.08$\pm0.01$&0.15$\pm0.03$&9.7$\pm0.9$\\
\hline
TDBC outside &0.18$\pm0.02$&0.10$\pm0.01$&0.10$\pm0.02$&10.8$\pm1.2$\\
\hline
RhB outside &0.18$\pm0.02$&0.21$\pm0.02$&0.20$\pm0.04$&11.7$\pm1.2$\\
\hline
\hline
\end{tabular}

\caption{\textbf{Spectral parameters for the different cavities}. Data were obtained as follows. Cavity mode widths were determined from the reflectance data with oscillators strengths for the dyes set to zero. Molecular widths were determined from modelling the transmission of thin films of the dyes using the Lorentz oscillator model and a transfer matrix approach. The Rabi splittings were estimated from the coupled oscillator models, the Finesse was determined from the mode width and the free spectral range.}
\label{tab:spectral_parameters}
\end{table}

\noindent\textbf{Optical Microcavities}:
The cavity mirrors were fabricated from gold using e-beam-assisted evaporation, with each mirror having a nominal thickness of 40 nm. PMMA with a molecular weight of 950K was spin-coated from a solution diluted in 4$\%$ Anisole to achieve a layer thickness of 120 nm. The microcavity was formed on a silicon substrate with a minimal oxide thickness layer.\\

\noindent\textbf{Material Parameters}:
For the TDBC/Rhodamine films we used as model are based on Lorentz oscillators,

\begin{equation}
\label{eq:Lorentz}
\varepsilon(\omega) = \varepsilon_{\infty}+ 
\sum_j^n\frac{\omega_{j}^2~f_j}{\omega_{j}^2-\omega^2-i\gamma_j\omega},
\end{equation}

\noindent with parameters given in table~\ref{tab:dl_parameters}. Note that we used a single oscillator model for TDBC and a two oscillator model for RhB, parameters for our RhB films were ascertained by ellipsometry.

\begin{table}[h!]
    \centering
     \begin{tabular}{| c| c| c| c |}
\hline 
  Molecules   & $f_j$ &$\omega_j$ (eV) & $\gamma_j$(eV)\\
 \hline
 \hline
 TDBC      &0.7 &  2.11 & 0.06  \\
 \hline
 RhB (j=1)    &0.08 & 2.21 & 0.14   \\
 \hline
 RhB (j=2)    &0.018 & 2.38 & 0.19   \\
 
 \hline
\end{tabular}
    \caption{Lorentz oscillator model parameters for different molecules types}
    \label{tab:dl_parameters}
\end{table}
\noindent and with $\varepsilon_{\infty}$ taken to be 1.99 for RhB and 1.90 for TDBC. For the Si substrate we took the permittivity to be -16+3i. For the gold we made use of a Drude model,

\begin{equation}
\label{eq:Drude-Lorentz}
\varepsilon(\omega) = \varepsilon_b- \frac{\omega_p^2}{\omega^2+i\gamma\omega},
\end{equation}

\noindent with parameters taken from Olmon {\it{et al.}} \cite{Olmon_PRB_2012_86_235147}, specifically, $\omega_p = 1.29\times 10^{16}~rad~s^{-1}$, and $\gamma = 7.30\times 10^{13}~rad~s^{-1}$, with  $\varepsilon_b = 1.0$.\\

\noindent\textbf{Coupled Oscillator Model}
Our coupled oscillator model can be expressed using the following matrix equation,

\begin{equation}
\begin{pmatrix}
\omega_{\textrm{cavity}}(k_x) & \Omega_{R}/2 \\ \Omega_{R}/2 & \omega_{\textrm{dye}} 
\end{pmatrix}
\begin{pmatrix}
a_{L1,\,U1}\\
b_{L1,\,U1}
\end{pmatrix}
= E_{L1,\,U1}
\begin{pmatrix}
a_{L1,\,U1}\\
b_{L1,\,U1}
\end{pmatrix},
\label{eqn:matrix_eqn_1}
\end{equation}

\noindent where $\omega_{\textrm{cavity}}(k_x)$ is the dispersion of the cavity mode, $\omega_{\textrm{dye}}$ is the resonance frequency of the dye. $\Omega_R/2$ is the coupling strength. $E_{L1,\,U1}$ are the eigenvalues of the lower and upper polaritons, and a,b are the Hopfield coefficients.

\twocolumngrid

\end{document}